\documentclass{osa-article}

\journal{osajournal}

\articletype{Research Article}

\begin{document}

\title{Real-time X-ray Phase-contrast Imaging Using SPINNet -- A Novel Speckle-based Phase-contrast Imaging Neural Network}

\author{Zhi Qiao,\authormark{*} Xianbo Shi,\authormark{**} Yudong Yao, Michael J. Wojcik, Luca Rebuffi, Mathew J. Cherukara, and Lahsen Assoufid}

\address{Advanced Photon Source, Argonne National Laboratory, Lemont, Illinois 60439, USA}

\email{\authormark{*}zqiao@anl.gov\\
\authormark{**}xshi@anl.gov} 



\begin{abstract}
X-ray phase-contrast imaging has become indispensable for visualizing samples with low absorption contrast. In this regard, speckle-based techniques have shown significant advantages in spatial resolution, phase sensitivity, and implementation flexibility compared with traditional methods. However, their computational cost has hindered their wider adoption. By exploiting the power of deep learning, we developed a new speckle-based phase-contrast imaging neural network (SPINNet) that boosts the phase retrieval speed by at least two orders of magnitude compared to existing methods. To achieve this performance, we combined SPINNet with a novel coded-mask-based technique, an enhanced version of the speckle-based method. Using this scheme, we demonstrate a simultaneous reconstruction of absorption and phase images on the order of 100 ms, where a traditional correlation-based analysis would take several minutes even with a cluster. In addition to significant improvement in speed, our experimental results show that the imaging resolution and phase retrieval quality of SPINNet outperform existing single-shot speckle-based methods. Furthermore, we successfully demonstrate its application in 3D X-ray phase-contrast tomography. Our result shows that SPINNet could enable many applications requiring high-resolution and fast data acquisition and processing, such as in-situ and in-operando 2D and 3D phase-contrast imaging and real-time at-wavelength metrology and wavefront sensing.
\end{abstract}

\section{Introduction}

With the development of high-brightness synchrotron facilities and free electron lasers, advanced X-ray imaging techniques including near-field phase-contrast imaging and far-field diffraction imaging are well-developed and widely used in material science, environmental science, and biomedical imaging~\cite{Park.2018.Popescu,Pfeiffer.2018.Pfeiffer,miao2011coherent,brown2002overview}. Compared with far-field diffraction-based methods, near-field phase-contrast imaging is a full-field imaging technique with lower coherence requirement, larger field of view, higher experimental flexibility, and can be performed with small-scale laboratory X-ray sources. In addition, phase-contrast X-ray imaging is more sensitive to sample density with low-Z materials than traditional transmission X-ray imaging, making it a preferred tool for studying low-absorption soft tissues such as muscles. 

During the past few decades, various phase-contrast imaging methods have been developed~\cite{Tao_2021_AppliedSciences, Bravin_2012_PhysicsinMedicineandBiology}, including the propagation-based method and grating interferometry, applied in 2D and 3D imaging of biomedical samples. However, quantitative information is missing from the propagation-based method \cite{paganin2018single,pavlov2020single} where extra constraints such as the sample's size and properties are required for the phase retrieval and reconstruction. Talbot grating interferometry can provide quantitative multi-contrast imaging, including absorption, phase, and dark-field~\cite{Inoue_2018_ReviewofScientificInstruments, VilaComamala_2021_OpticsExpress, Zdora_2017_BiomedicalOpticsExpress,Assoufid_2016_ReviewofScientificInstruments}, but suffers from complicated setup, high source coherence requirement, phase warping, and relatively low spatial resolution. 

Recently, the speckle-based imaging (SBI) method has been applied to quantitative X-ray multi-contrast imaging, at-wavelength metrology, and wavefront sensing~\cite{Zdora_2017_PhysicalReviewLetters, Zdora_2020_Optica,Zdora_2018_JournalofImaging, Berujon_2016_PhysicalReviewApplied}. Compared with Talbot grating interferometry, the SBI method's primary advantages are spatial resolution, phase sensitivity, and experimental flexibility~\cite{Zdora_2018_JournalofImaging, Zdora_2020_Optica,Wang_2013_JournalofPhysicsConferenceSeries}. Most SBI schemes, including single-shot X-ray speckle tracking (XST), X-ray speckle-scanning tracking (XSS), and X-ray speckle-vector tracking (XSVT), mainly use digital image correlation (DIC) to track the speckle movement~\cite{Berujon_2016_PhysicalReviewApplied,Berujon2017near,Wang2016quantitative}. An optimization-based method called unified modulated pattern analysis (UMPA) was later proposed to deal with broader situations, including different scan positions, spatial resolution, and phase sensitivity for either random or periodic patterns~\cite{Zdora_2017_PhysicalReviewLetters}. Most recently, a coded-mask-based multi-contrast imaging (CMMI) technique has been developed~\cite{Qiao2021single}, which uses a coded phase mask to generate a pre-known pattern with ultra-high-contrast. Combined with the advanced optimization method, CMMI has shown superior performance in spatial resolution and phase sensitivity compared with existing methods. However, the computational efficiency of current SBI methods, whether DIC-based or optimization-based, is a major obstacle for their application in in-situ and in-operando measurements, where real-time analysis is critical. Data processing using the current correlation-based analysis methods may take several minutes even with a cluster, even when high-performance computing resources are employed. The SBI method using multi-resolution analysis and wavelet-transform has significantly improved the data processing speed~\cite{Qiao2020multiresolution,Qiao2020wavelet}. However, the improvement is still far from the requirement of many in-situ measurements. In addition, with the advent of next-generation synchrotron facilities, such as the Advanced Photon Source Upgrade project~\cite{borland2018upgrade}, the experimental data volume and data acquisition are expected to be many orders of magnitudes larger and faster compared with the current state. Therefore, high-speed image processing is essential and will open many new opportunities in real-time applications such as in-situ multi-contrast imaging~\cite{liu2018high} and real-time wavefront sensing and control~\cite{leung2018situ}. 

Machine learning has been actively investigated in computational imaging methods to achieve super-resolution, low-flux imaging, and high-speed imaging~\cite{Lee_2020_ScientificReports,Wang_2020_ScientificReports,Rivenson_2017_LightScienceApplications,Rivenson_2019_LightScienceApplications, Li_2018_Optica, Goy_2018_arXiv,Jo_2018_IEEEJournalofSelectedTopicsinQuantumElectronics, Sinha_2017_Optica}. Inspired by the deep-learning-based optical flow methods~\cite{Sun_2017_arXiv,Fischer_2015_arXiv,Hui_2018_arXiv,Hui_2019_arXiv,Hui_2020_arXiv,Liu_2019_arXiv} for object detection in the computer vision field, we propose a novel speckle-based phase-contrast imaging neural network (SPINNet) for single-shot imaging. It improves the data processing speed by more than two orders of magnitudes compared to existing X-ray speckle-based phase-contrast imaging methods. Using SPINNet, we demonstrate simultaneous phase and amplitude recovery within 100 ms. In addition, the image quality also outperforms the DIC-based XST method as quantified by the noise level. Our results show that SPINNet could enable in-situ and in-operando SBI measurements in 2D and 3D phase-contrast imaging.

\section{Methods and Network}

\subsection{Experimental setup and principle}

\begin{figure}[ht!]
	\centering\includegraphics[width=0.7\textwidth]{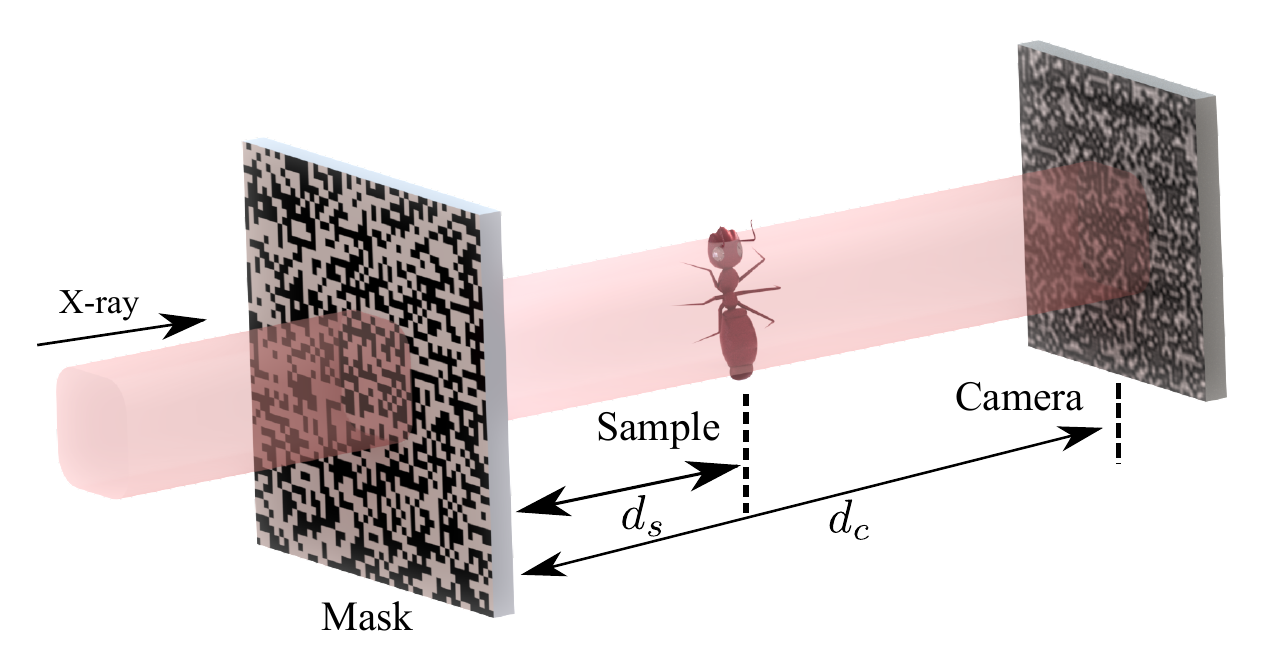}
	\caption{Schematic of a typical speckle-based phase-contrast imaging setup with a collimated beam. Note that the coded mask can be replaced by any speckle generator. $d_s$ and $d_c$ are the mask-to-sample, mask-to-camera distances, respectively.}
	\label{fig:setup}
\end{figure}

The schematic of the experimental setup (also a typical SBI setup) is shown in Fig.~\ref{fig:setup}, where the X-ray beam passes through the speckle generator (e.g., a coded mask in case of CMMI), the sample at a mask-to-sample distance, $d_s$, and then impinges on the detector camera at a mask-to-camera distance, $d_c$.  The speckle pattern generated by the coded mask will be distorted by the sample transmission $T (x,y)$ and phase $\phi (x,y)$ distribution. Based on the near-field propagation process and the small-angle approximation, the sample-in image (referred to as sample image hereafter),  $I_s(x, y)$, on the detector camera can be expressed as,
\begin{equation}\label{eq:propagation}
\begin{split}
	I_s(x, y) &= T(x,y) \times I_r(x - \delta x, y - \delta y),\\
	\delta x (x, y) &= \frac{\lambda (d_c-d_s)}{2\pi}\frac{\partial\phi (x, y)}{\partial x},\\
	\delta y (x, y) &= \frac{\lambda (d_c-d_s)}{2\pi}\frac{\partial\phi (x, y)}{\partial y},
\end{split}
\end{equation}
where $I_r(x,y)$ is the reference speckle image without the sample in the X-ray beam (refer to as reference image hereafter) and $\lambda$ is the X-ray wavelength. Hereafter, we omit the coordinate $(x, y)$ for simplicity. The speckle tracking process is thus to find the local pixel-wise displacements ($\delta{x}$ and $\delta{y}$) from the reference image ($I_r$) to the sample image ($I_s$). From the displacement maps ($\delta{x}$ and $\delta{y}$), differential phase images ($\partial \phi / \partial x$ and $\partial \phi / \partial y$) can be extracted based on Eq.~(\ref{eq:propagation}). Then the sample phase $\phi$ can be reconstructed by the 2D integration of differential phase $\partial \phi / \partial x$ and $\partial \phi / \partial y$. 

The experiment was carried out at the 1-BM beamline of the Advanced Photon Source (APS). The X-ray photon energy was set to 14~keV using a Si(111) double-crystal monochromator. To obtain a high-contrast speckle pattern, a coded binary phase mask with a pitch size of 5~\textmu m was used as the speckle generator \cite{Qiao2021single}. The coded mask was fabricated by electroplating 2~\textmu m thick Au into polymer molds patterned using electron beam lithography (see \textcolor{urlblue}{Supplement 1} for details on the coded-mask pattern). The detector camera system had an effective pixel size of 0.65~\textmu m, which consisted of a 100~\textmu m thick LuAG:Ce scintillator, a 10$\times$ objective, a 45\textdegree~reflective mirror, and an Andor Neo sCMOS camera. The sample-to-camera distance was $d_c - d_s = $ 628~mm. For the 2D imaging experiment, only one speckle image pair with one reference image and one sample image was recorded. For the 3D tomography measurement, the reference image was acquired only once before putting the sample in the beam,  while the sample images were collected at each sample rotation angle. 

\subsection{SPINNet design structure} \label{sec:net_structure}

\begin{figure}[ht!]
	\centering\includegraphics[width=1\textwidth]{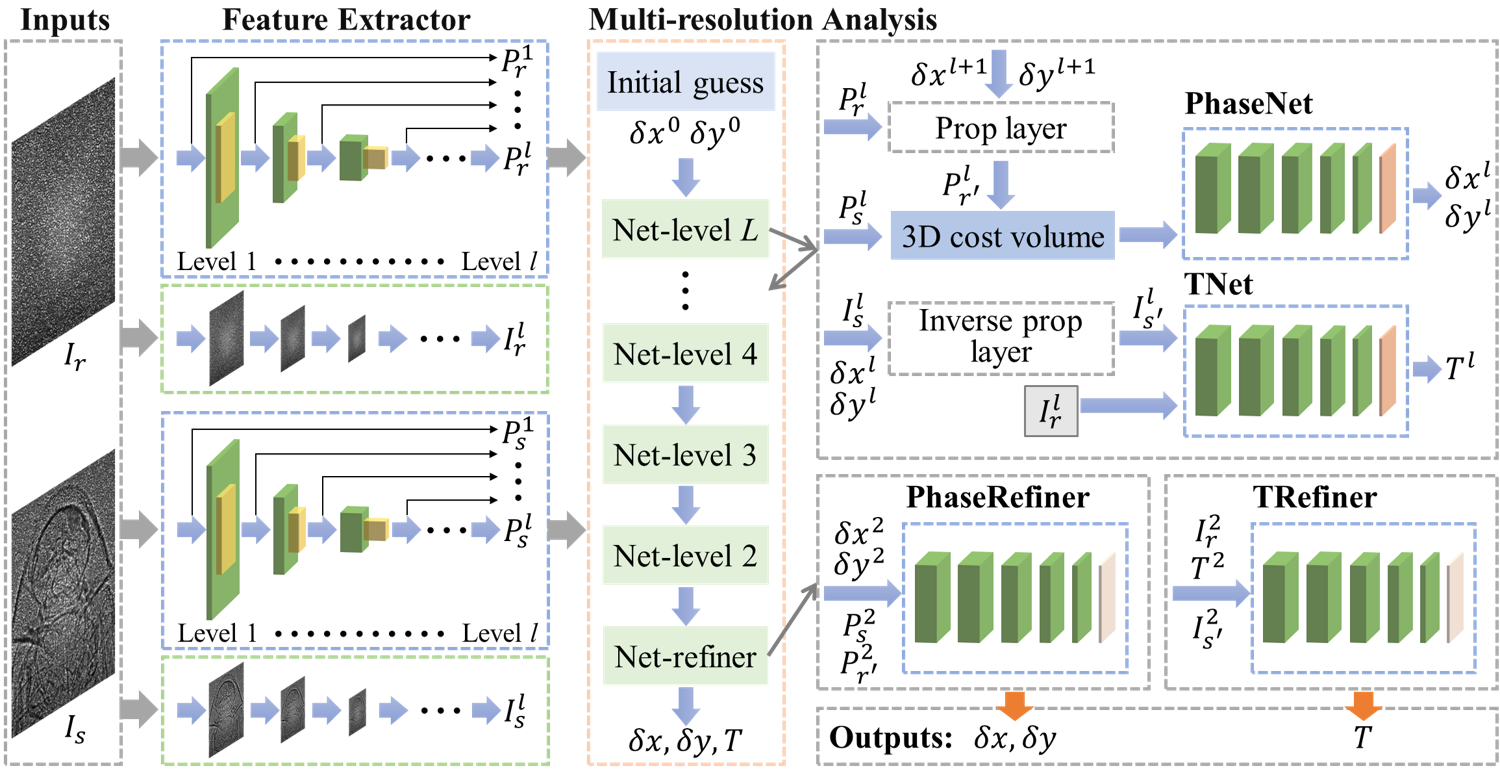}
	\caption{Schematic of SPINNet structure.}
	\label{fig:network}
\end{figure}

The design structure of SPINNet is based on the optical flow networks, which have been used for object detection in self-driving~\cite{Hui_2018_arXiv,Sun_2017_arXiv}. As shown in Fig.~\ref{fig:network}, SPINNet consists of three sub-networks: the feature extractor for reference and sample images, the estimator (PhaseNet and TNet for phase and transmission, respectively), and the refiner (PhaseRefiner and TRefiner for phase and transmission, respectively). The detailed structures for each sub-network can be found in \textcolor{urlblue}{Supplement 1}. The inputs for SPINNet are one reference image ($I_r$) and one sample image ($I_s$), while the outputs are the transmission image ($T$) and two displacement maps ($\delta x$ and $\delta y$). 

The SPINNet structure also includes the multi-resolution analysis and 3D cost volume evaluation, essential features for speckle tracking to achieve the best efficiency and accuracy. The multi-resolution analysis has been demonstrated to improve the calculation efficiency of the conventional correlation-based algorithms by reducing the adequate searching window size in each of the multi-resolution levels~\cite{Qiao2020multiresolution}. It is particularly effective for samples with large phase gradients and coexistent fine features. Here we introduce the multi-resolution analysis into SPINNet in the feature extractor and estimator sub-networks. The effects of different resolution levels can be found in Fig.~S3 in \textcolor{urlblue}{Supplement 1}. The 3D cost volume is another key feature of SPINNet to account for the process of sub-window correlation, which is an extra but necessary physical constraint. 

The multi-resolution process starts with the feature extractor, which extracts feature information of different domain sizes from the input reference image ($I_r$) and sample image ($I_s$). Since the reference and sample images can have different features, two separate feature extractors are built for $I_r$ and $I_s$, respectively. A max multi-resolution level of $L=6$ is used in the current SPINNet. For each resolution level, the image size will be down-sampled from a lower level by a factor of 2 in each image coordinate dimension ($x$ and $y$) and, at the same time, the feature information $P^l$ will have an increasing channel number $K$ as the third dimension. Details of the extractor functions and the dimensions of $P^l$ can be found in \textcolor{urlblue}{Supplement 1}. 

The estimator contains multi-resolution levels (named as Net-level $l$, with $l \in [2,L]$) as shown in Fig.~\ref{fig:network}. There are totally $(L-1)$ PhaseNets and $(L-1)$ TNets for image features $P_r^l$, $P_s^l$, $I_r^l$ and $I_s^l$ with different sampling resolution. 

Firstly, the initial displacement ($\delta x^0$ and $\delta y^0$) and transmission ($T^0$) maps are generated randomly for the highest multi-resolution level (Net-level $L$). The reference image feature $P_r^l$ is propagated to sample image position using the displacement maps ($\delta x^{l+1}$ and $\delta y^{l+1}$) from the higher level for each feature channel ($k \in [1,K]$) according to Eq.~(\ref{eq:propagation}). This propagation process is called the prop-layer, denoted as
\begin{equation}\label{eq:p_layer}
	P^l_{r'}(x, y, k^l) = P^{l}_r(x - \delta x^{l+1}, y - \delta y^{l-1}, k^l).
\end{equation}
In this step, we set the transmission $T=1$ to reduce the cross-talk between the phase shift and transmission in the neural network. Then the 3D cost volume is constructed as
\begin{equation}\label{eq:cost_vol}
	C(x, y, h) = \frac{\sum_k{P^l_{r'}(x-p, y-q, k)P^l_s(x, y, k)}}{K},
\end{equation}
where $h=p\times (2N+1) + q$, and $p, q \in (-N, N)$ are distances of nearby pixels within the max searching range $N$ along $x$ and $y$, respectively. The total channel dimension of the 3D cost volume $C$ in the $h$ coordinate is $(2N+1)^2$. Thanks to the multi-resolution approach, a search range of $N=3$ or 4 is adequate in each level, giving a channel dimension of $C$ to be 49 or 81, respectively. Once the 3D cost volume is built, the displacement $\delta^l{x}$ and $\delta^l{y}$ for the multi-resolution level $l$ can be obtained by passing through the PhaseNet. $\delta^l{x}$ and $\delta^l{y}$ also become the input for the next lower level estimator.

Unlike the phase prediction, the multi-resolution reference image $I^l_r$ and sample image $I^l_s$ without going through feature extractors are used for the transmission prediction because the relative intensity information might be missing in the image feature $P_r^l$ and $P_s^l$. $I_r^l$ and $I_s^l$ are obtained simply by image downsizing using bilinear interpolation, by a factor of 2 in $x$ and $y$ for each level.The inverse prop-layer, which is an inverse process of Eq.~(\ref{eq:p_layer}), is used to propagate the sample image ($I_s^l$) to the reference image domain ($I_{s'}^l$). Then TNet will predict the transmission image $T^l$ based on $I_r^l$ and $I_{s'}^l$ at each resolution level $l$. 

The above procedure will be repeated to reach the last multi-resolution level (typically $l = 2$ or 3). Finally, a PhaseRefiner and a TRefiner are added to remove the extra noise and improve the accuracy of the predicted displacement and transmission images. All subnets (PhaseNet, TNet, PhaseRefiner, and TRefiner) use similar conventional neural network structures described in detail in Fig.~S2 in \textcolor{urlblue}{Supplement 1}. Here a multi-resolution level of $l = 2$ instead of $l = 1 $ is used for faster training and prediction speed by reducing the network size. From experience, we note that no obvious improvement can be observed by introducing the level of $l = 1$. Considering that $P_r^1$ and $P_s^1$ are original reference image ($I_r$) and sample image ($I_s$), respectively, without passing through the feature extractors, which contains no feature information, it is reasonable to discard the multi-resolution level of $l = 1$.

Once the displacements $\delta x$ and $\delta y$ and the transmission $T$ are obtained by SPINNet, the following loss function is used to characterize the prediction error,
\begin{equation}\label{eq:cost_func}
	L = \frac{||\delta{x} - \delta_{gt}{x}||_2^2}{||\delta_{gt}{x}||_2^2} + \frac{||\delta{y} - \delta_{gt}{y}||_2^2}{||\delta_{gt}{y}||_2^2} + \frac{||T - T_{gt}||_2^2}{||T_{gt}||_2^2},
\end{equation}
where $||\cdot||_2$ represents $L^2$ norm, $\delta_{gt}{x}$, $\delta_{gt}{y}$ and $T_{gt}$ are ground-truth images of the displacements and transmission, respectively.

\subsection{Training data generation} \label{sec:trainingdata}

The training data generation is critical for a neural network and can affect the accuracy and general applicability in analyzing actual experimental data. Below we discuss how to generate the reference and sample image pairs based on Eq.~(\ref{eq:propagation}) to best approximate the experimental conditions.

\subsubsection{Reference image generation}
In the conventional speckle-tracking methods, sandpaper or filter membrane is usually used as the speckle generator, which is inexpensive and straightforward to implement. However, the random speckle pattern has various parameters such as speckle feature size, blurring effect, and contrast to add extra complexity to the neural network training. Thus, numerous training data points will be needed to cover different experimental conditions. On the other hand, the coded mask was proposed recently to generate speckle patterns with ultra-high contrast~\cite{Qiao2021single}. The pre-knowledge of coded-mask design parameters significantly reduces the data volume required for network training. Therefore, we choose to generate the simulation data based on the coded binary phase mask.

\begin{figure}[ht!]
	\centering\includegraphics[width=0.68\textwidth]{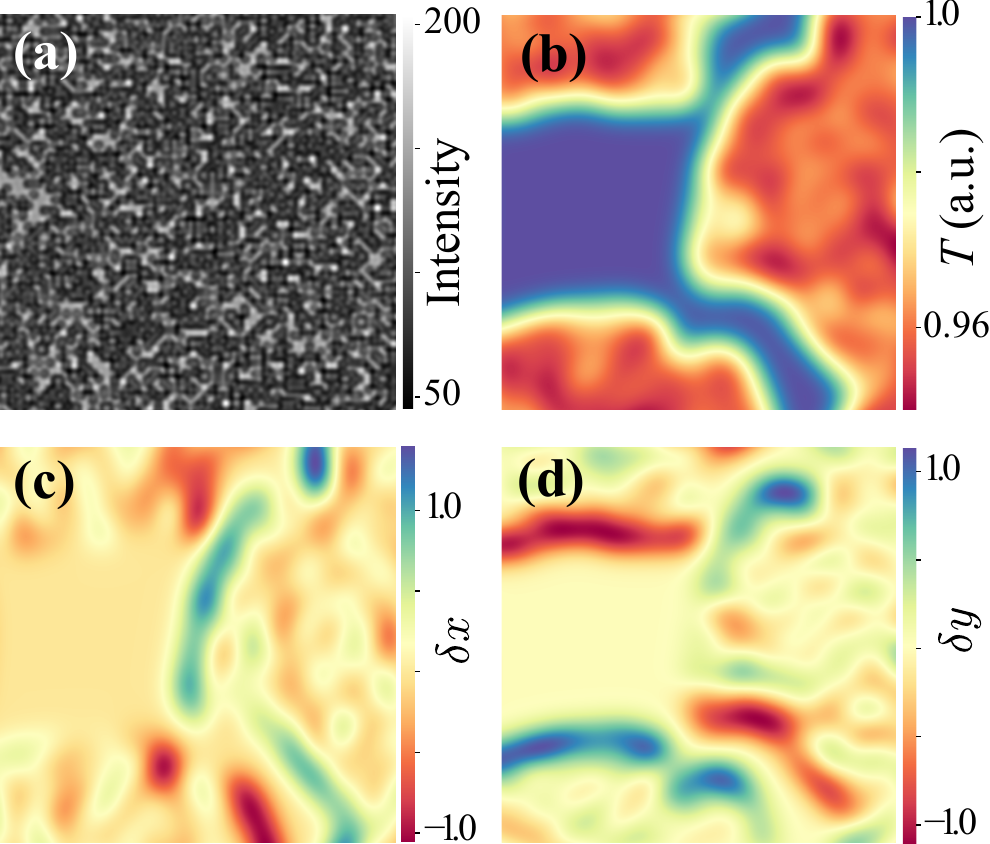}
	\caption{Examples of simulated training data for SPINNet: (a) reference image, (b) transmission image, (c) horizontal displacement, $\delta{x}$, and (d) vertical displacement, $\delta{y}$.}
	\label{fig:simu}
\end{figure}

Each reference image is generated by propagating a plane wave modulated by a randomly generated binary-phase-mask pattern. We consider all coded-mask patterns to have a total size of $M\times M$ pixels with a mask pitch size of $n$ pixels. Firstly, a random binary noise image with a size of $M/n\times M/n$ pixels is generated to have half pixels equal to 0 and the other half equal to 1. Then this binary noise image is upsampled by a factor of $n$ without interpolation to produce an image $R$ with the size of $M\times M$ pixels to represent the pattern of a coded mask. The reference image is obtained by propagating a plane wave with its amplitude and phase modulated by $R$ to a distance $d$ using the Fresnel diffraction formula,
\begin{equation}\label{eq:diff_nf}
	I_r=\left|\frac{e^{jkd}}{j\lambda d}\iint (A_0 + t_0 R)e^{j\phi_0 R}e^{jk\frac{(x-x_0)^2+(y-y_0)^2}{2d}}dx_0dy_0\right|^2,
\end{equation}
where $k=2\pi/\lambda$, $A_0$ and $t_0$ define the wavefront amplitude and the consequent image intensity and contrast, and $\phi_0$ is the phase shift of the coded mask. At last, the reference image $I_r$ is Gaussian filtered with a kernel size of 3 pixels, which mimics the limited resolution of the detector system. We set the simulation parameters to $\lambda=0.06$~nm, $d=0.3$~m, $A_0=0.7$, $t_0=0.3$, and $\phi_0=\pi$. Note that these parameters do not need to be the same as the real condition, as long as the produced pattern images have similar feature sizes and contrast as the experimental data. Figure.~\ref{fig:simu}(a) shows an example of a reference image.

\subsubsection{Sample image generation}

A sample image is generated using one of the above reference images and  Eq.~(\ref{eq:propagation}) with the displacement ($\delta{x}$ and $\delta{y}$) and transmission ($T$) maps created randomly as described below.

Firstly, a Gaussian noise image ($M\times M$ pixels) is generated with a mean value of 0.5, normalized to the range of [0, 1], and Gaussian filtered with a kernel size of 10 pixels to remove high-frequency noises. Note that the smallest sample feature size can be determined by choosing the Gaussian filter kernel size carefully. The Bézier curve is then used to create a sample contour image. The sample phase image is obtained by multiplying the noise image with the sample shape image and then scaled by a factor $V_p$ randomly generated within a range of $[\pi, 20\pi]$ for each image. Once the sample phase image is generated, the displacement maps ($\delta{x}$ and $\delta{y}$) are calculated as the image gradients along $x$ and $y$ directions, respectively, following Eq.~(\ref{eq:propagation}). 

The generation of sample transmission $T$ is similar to the phase generation process, except that the transmission image is scaled by a factor $V_T$ randomly selected in a range of $[0.02, 0.2]$ for each image. Considering that the sample transmission image normally has the same shape and features as its phase image, we set 40\% of the transmission images to have the same distribution as the phase image, and the rest 60\% have independent distributions. Example images of sample transmission $T$, horizontal displacement $\delta{x}$, and vertical displacement $\delta{y}$ are shown in Figs.~\ref{fig:simu}(b), (c) and (d), respectively.

\subsection{SPINNet training} \label{sec:training}

\begin{figure}[ht!]
	\centering\includegraphics[width=0.68\textwidth]{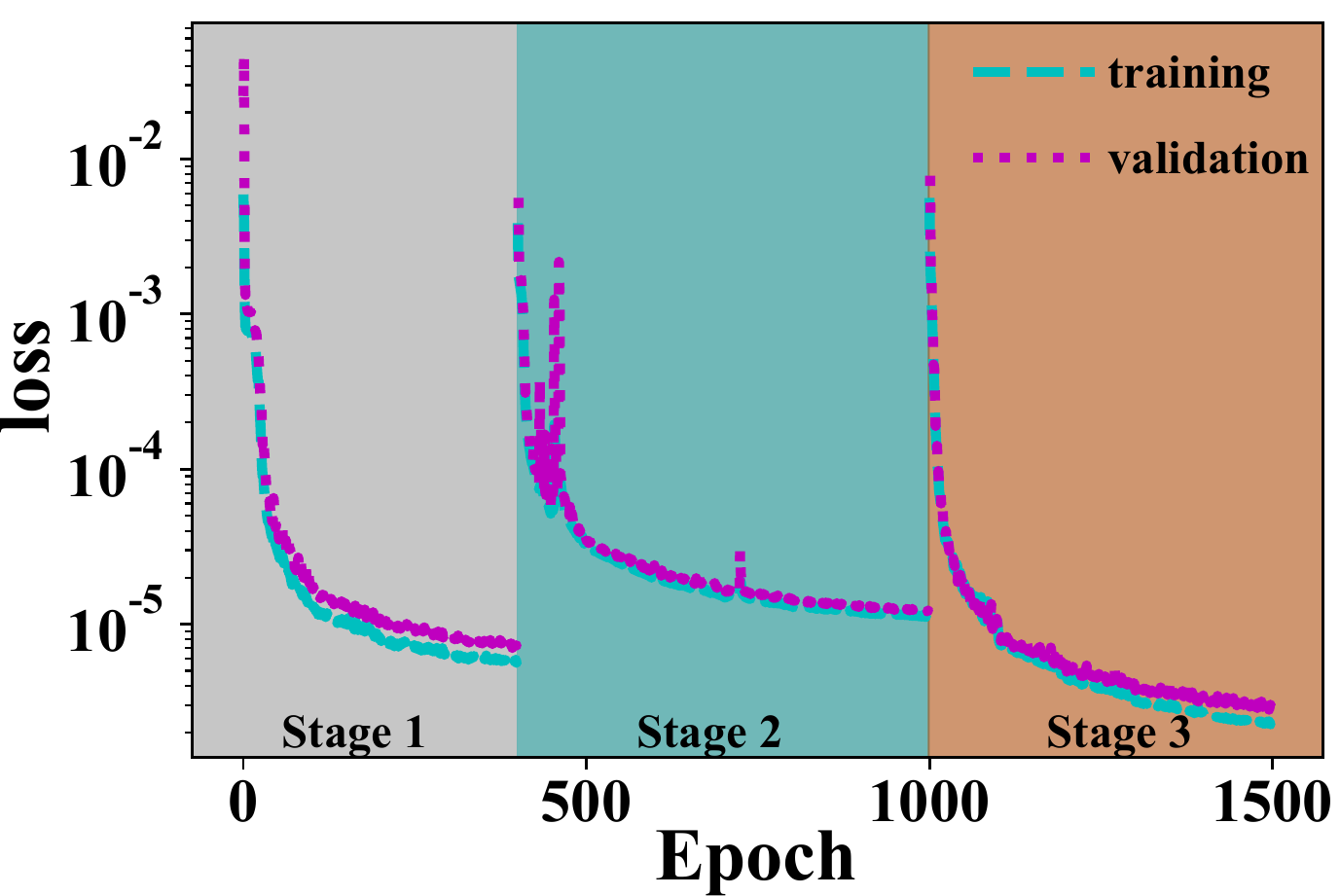}
	\caption{Training and validation loss as a function of epoch within three stages. Stage 1 (epoch 1\textasciitilde 400) is for Net-level 6\textasciitilde3. Stage 2 (epoch 401 \textasciitilde 1000) is for Net-level 6\textasciitilde2. Stage 3 (epoch 1001 \textasciitilde 1500) is for Net-level 6\textasciitilde2 with refiners.}
	\label{fig:train_loss}
\end{figure}

A dataset of 10,000 simulated reference-sample image pairs is generated with 80\% for training and 20\% for validation. SPINNet is trained within three stages from the low-resolution to the high-resolution level using the Adam optimizer with an initial learning rate of 1.0E-4. The first stage trains the Net-level 6 to Net-level 3 within 400 epochs until convergence. In the second stage, Net-level 2, initialized with the weight from the former Net-level 3, is added into the network and trained within 600 epochs. The third stage trains the whole network with the PhaseRefiner and TRefiner within 500 epochs. For a better convergence, the learning rate is reduced by half for every 100 epochs. The number of parameters for SPINNet is around 5 million. The whole training takes 8 hours using 16 NVIDIA A100 GPUs with a batch size of 96, 45, and 8 for the three stages. The training and validation loss based on Eq.~(\ref{eq:cost_func}) is shown in Fig.~\ref{fig:train_loss}. The use of a refiner effectively reduces the training loss in stage 3. The validation loss and training loss are almost the same, indicating no obvious overfitting during training.

\section{Results and Analysis}
\subsection{Simulation data reconstruction} \label{sec:simu_infer}

\begin{figure}[ht!]
	\centering\includegraphics[width=0.78\textwidth]{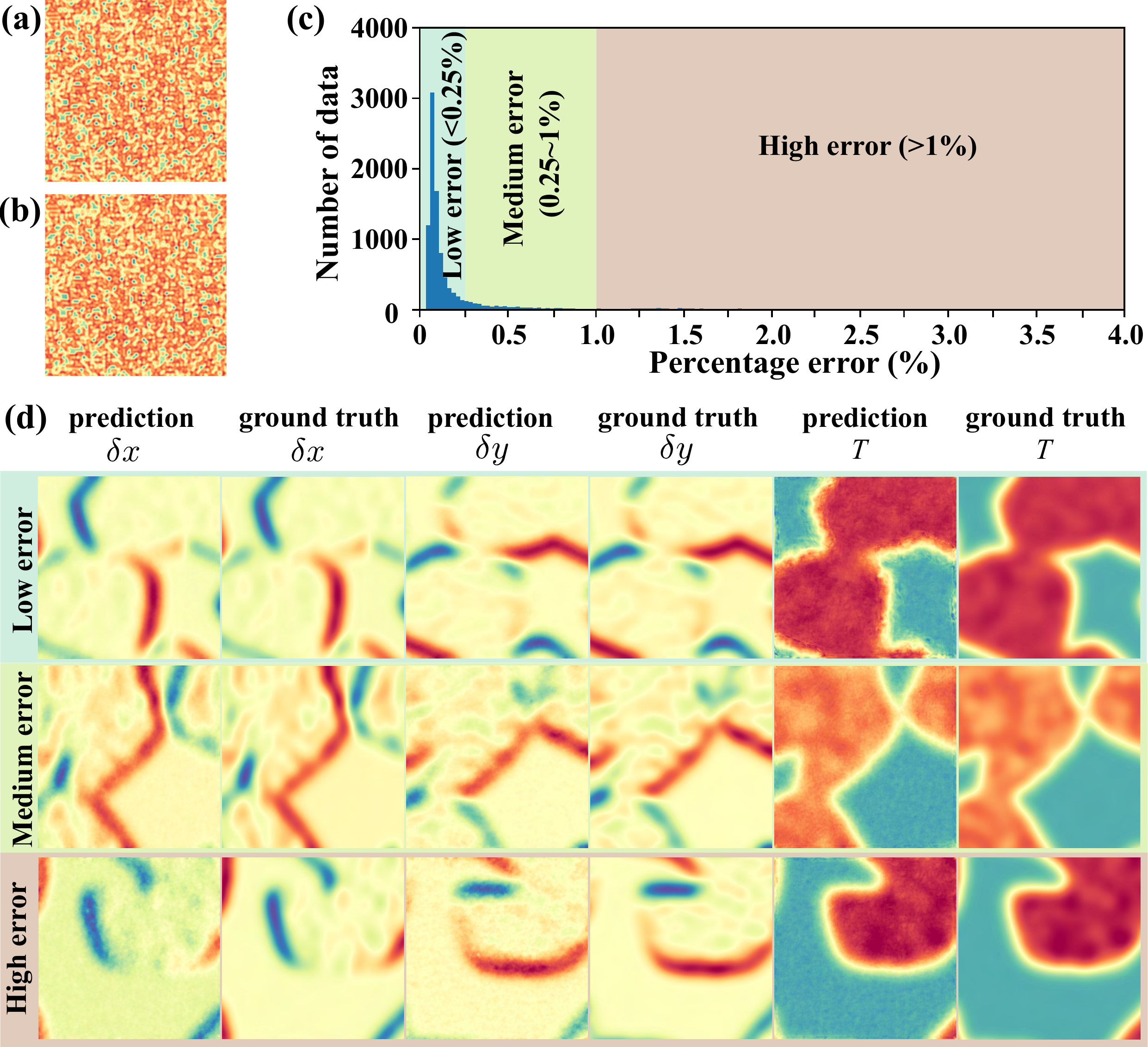}
	\caption{Simulated (a) reference image and (b) sample image. (c) Histogram of the prediction error distribution of a 10,000 test dataset using SPINNet. Example prediction results and their ground truth from each error regime. }
	\label{fig:inference_statis}
\end{figure}

We first use 10,000 separately simulated data to evaluate the performance of SPINNet. Figures~\ref{fig:inference_statis}(a) and (b) show a simulated reference and sample images as an example. The histogram in Fig.~\ref{fig:inference_statis}(c) shows the statistics of the percentage error between the SPINNet prediction and the ground truth for all 10,000 data. We define a low error regime with a percentage error of less than 0.25\%, a medium error regime with 0.25\%\textasciitilde1.0\%, and a high regime with a larger than 1\% error. There are only 7\% of data that have prediction errors above 1\%, but all are still less than 4\%. Examples of the predicted displacements and transmission and their ground truth from each error regime are also shown in Fig.~\ref{fig:inference_statis}(d). Even in the high error regime, the displacement prediction agrees well with the ground truth regarding low-frequency features and with only slight difficulty in predicting high-frequency components.  Figure~\ref{fig:inference_statis}(d) also indicates that the displacement (or phase) prediction has less noise than the transmission prediction because PhaseNet is based on 3D cost volume with a search range of 3 pixels, so that an extra physical constraint is implemented. At the same time, TNet predicts the transmission in a pixel-wise fashion without any constraint. Although the prediction of transmission has more noise, the overall distribution and structure are all accurately reconstructed. 

\subsection{Experimental data reconstruction}
In this section, we test the SPINNet performance with experimental data and compare the predicted results with the conventional DIC-based XST analysis. The SPINNet reconstruction was carried out using one NVIDIA A100 GPU, while the DIC-based XST analysis was performed on a cluster with 24-core Xeon E5-2670 CPU and 32 GB RAM. The DIC-based XST analysis used a template window size of 7$\times$7 and a searching window size of 20$\times$20 pixels. 

\subsubsection{Quantitative phase reconstruction}

The reconstruction accuracy is essential for applications requiring quantitative information such as X-ray at-wavelength metrology and quantitative phase-contrast imaging. For X-ray at-wavelength metrology, state-of-the-art optics such as refractive lenses and total-reflection mirrors have figure errors that require a phase measurement accuracy in the $\lambda /100$ level \cite{xianbo_optics}. Here we test the SPINNet reconstruction accuracy by measuring a Beryllium X-ray lens and comparing the results with conventional DIC-based XST analysis.

\begin{figure}[ht!]
	\centering\includegraphics[width=0.78\textwidth]{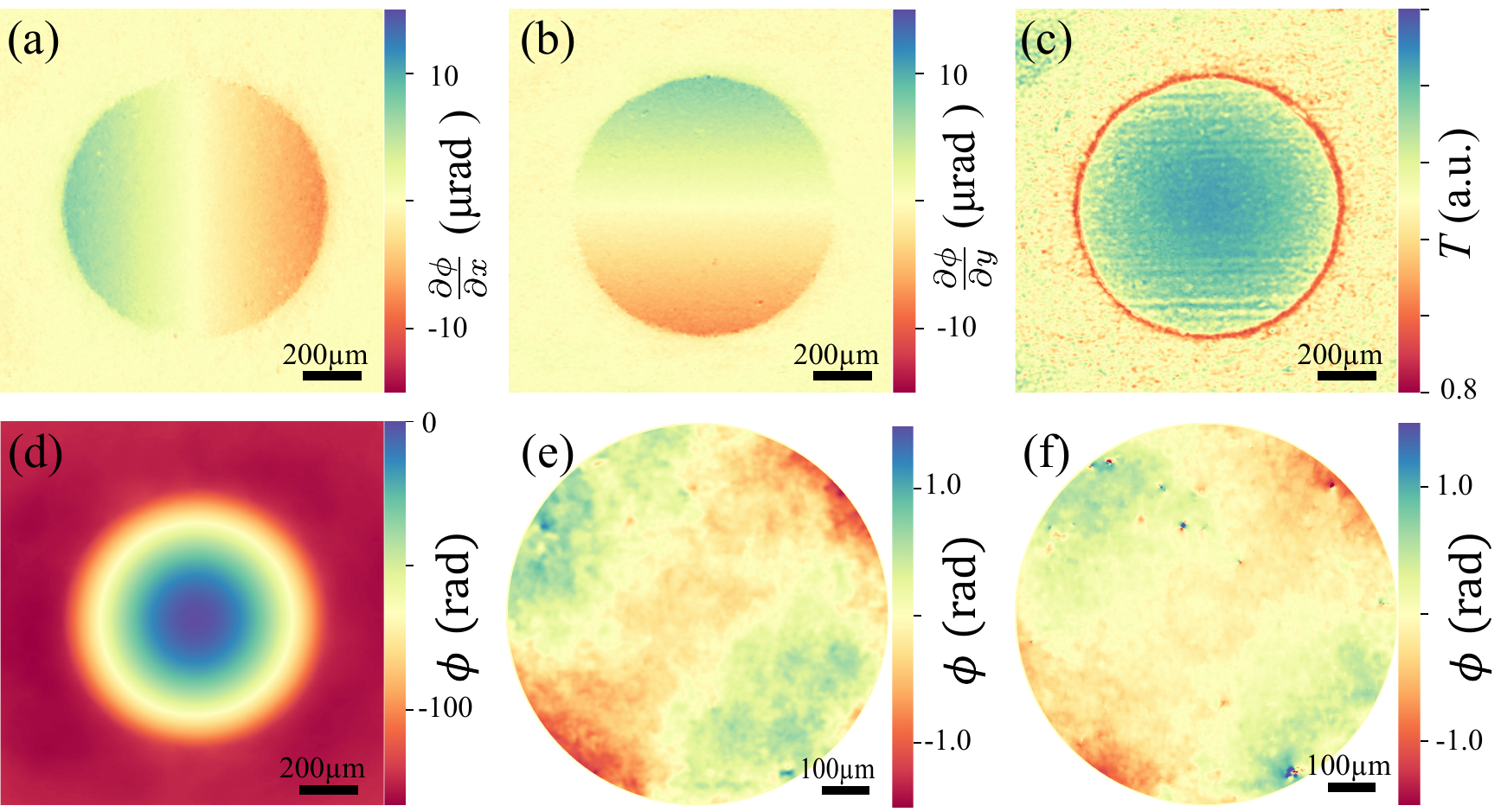}
	\caption{SPINNet reconstructed (a) horizontal differential phase, (b) vertical differential phase, (c) transmission image, (d) phase profile, and (e) residual phase error after removing the best fit parabola from (d) of a 2D parabolic beryllium lens. (f) Residual phase error result using DIC-based XST analysis for the same lens.}
	\label{fig:CRL_compare}
\end{figure}

A 2D parabolic beryllium lens with an apex radius of 200~\textmu m was measured using the single-shot speckle-tracking setup in Fig.~\ref{fig:setup} with a sample-camera distance of 500~mm. SPINNet predicted the horizontal differential phase, vertical differential phase, transmission, and integrated phase profiles as shown in Figs.~\ref{fig:CRL_compare}(a)-(d) using the measured reference and sample images. After subtracting the best fit parabola from the phase profile in Fig.~\ref{fig:CRL_compare}(d), the residual phase error within a circular area of 800~\textmu m diameter was obtained and shown in Fig.~\ref{fig:CRL_compare}(e). For comparison, the lens phase was also reconstructed using the DIC-based XST analysis, and the resulting residual phase error is shown in Fig.~\ref{fig:CRL_compare}(f). The relative rms error between the residual phase in Figs.~\ref{fig:CRL_compare}(e) and (f) is about 0.15 rad (0.02~$\lambda$), which is close to the theoretical phase sensitivity (0.01~$\lambda$) of the experimental setup.

Regrading the data processing speed, the DIC-based XST analysis took several minutes. In contrast, the SPINNet prediction took only 0.16~second with a batch size of 1, which can be further reduced by increasing the batch size: for example,  0.13~second computation time is obtained with a batch size of 3. The ultra-fast speed and high accuracy prove the feasibility of SPINNet in real-time X-ray wavefront sensing and beamline diagnostics at synchrotron light sources and free-electron lasers beamlines.

\subsubsection{Phase-contrast imaging comparison} \label{sec:exp_compare}

\begin{figure}[ht!]
	\centering\includegraphics[width=0.88\textwidth]{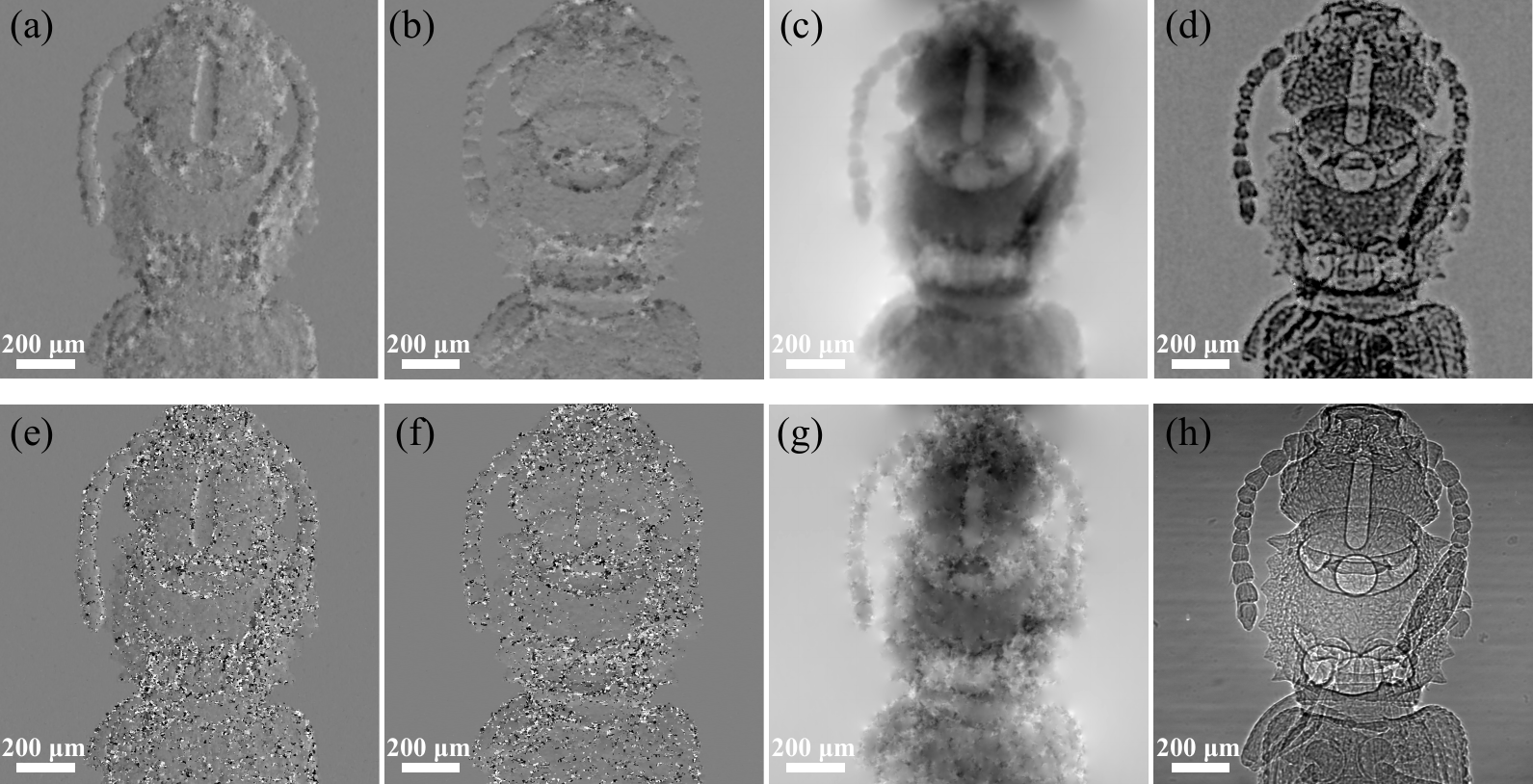}
	\caption{Comparison of phase reconstruction results using SPINNet and DIC-based XST analysis of a flour bug. (a) Horizontal differential phase, (b) vertical differential phase, (c) phase, and (d) transmission image reconstructed by SPINNet. (e) Horizontal differential phase, (f) vertical differential phase, and (g) phase image reconstructed by the DIC-based XST analysis. (h) Raw radiography image.}
	\label{fig:exp_compare}
\end{figure}

A flour bug was measured using the same setup in Fig.~\ref{fig:setup}. The reconstructed results using SPINNet are shown in Fig.~\ref{fig:exp_compare} and compared with results using DIC-based XST analysis. The differential phase images using SPINNet [Fig.~\ref{fig:exp_compare}(a) and (b)] have much lower noise than those using the DIC-based XST analysis [Fig.~\ref{fig:exp_compare}(e) and (g)]. As a result, the phase image using SPINNet [Fig.~\ref{fig:exp_compare}(c)] has significantly better quality and higher spatial resolution than that of the DIC-based XST analysis [Fig.~\ref{fig:exp_compare}(g)]. Also, the SPINNet predicted transmission image [Fig.~\ref{fig:exp_compare}(d)] shows identical sample features as in the radiography image [Fig.~\ref{fig:exp_compare}(h)], but with much higher contrast. 

Because of the optimized network structure, the SPINNet reconstruction outperforms the DIC-based XST analysis in image quality. Two main factors contribute to the superiority of SPINNet: (1) the network is based on convolution layers acting as low-pass filters to suppress image noise, and (2) SPINNet is not only based on the image pixel value, but also extracted features using feature extractors. In addition, the multi-resolution process takes into account the feature information with pyramid resolution levels, which may also help the image reconstruction quality. 

For the DIC-based XST analysis, the template window size directly limits the available spatial resolution. A small template window size gives rise to a better spatial resolution at the expense of a larger noise level. Instead, SPINNet predicts phase and transmission in a pixel-wise fashion based on the feature information and the 3D cost volume to obtain a higher spatial resolution.

The high speed of SPINNet is also the result of the multi-resolution process and thus the reduced search window size for the 3D cost volume. The searching window size of the 3D cost volume is only 3$\times$3 pixels in SPINNet, while that of the DIC-based XST analysis is 20$\times$20 pixels.

It should also be noted that even SPINNet was trained with simulated data using code mask patterns of 5~\textmu m pitch size, the trained model also works well for experimental images with a different mask pitch size. Prediction results of the same flour bug with experimental data obtained using a coded mask of 2~\textmu m pitch in Fig.~S4 (see \textcolor{urlblue}{Supplement 1}) show high image quality and resolution with only slightly higher noise. Therefore, SPINNet is not sensitive to the training data and can be a general tool for speckle-based phase-contrast imaging.

\subsubsection{Phase-contrast tomography}

\begin{figure}[ht!]
	\centering\includegraphics[width=0.88\textwidth]{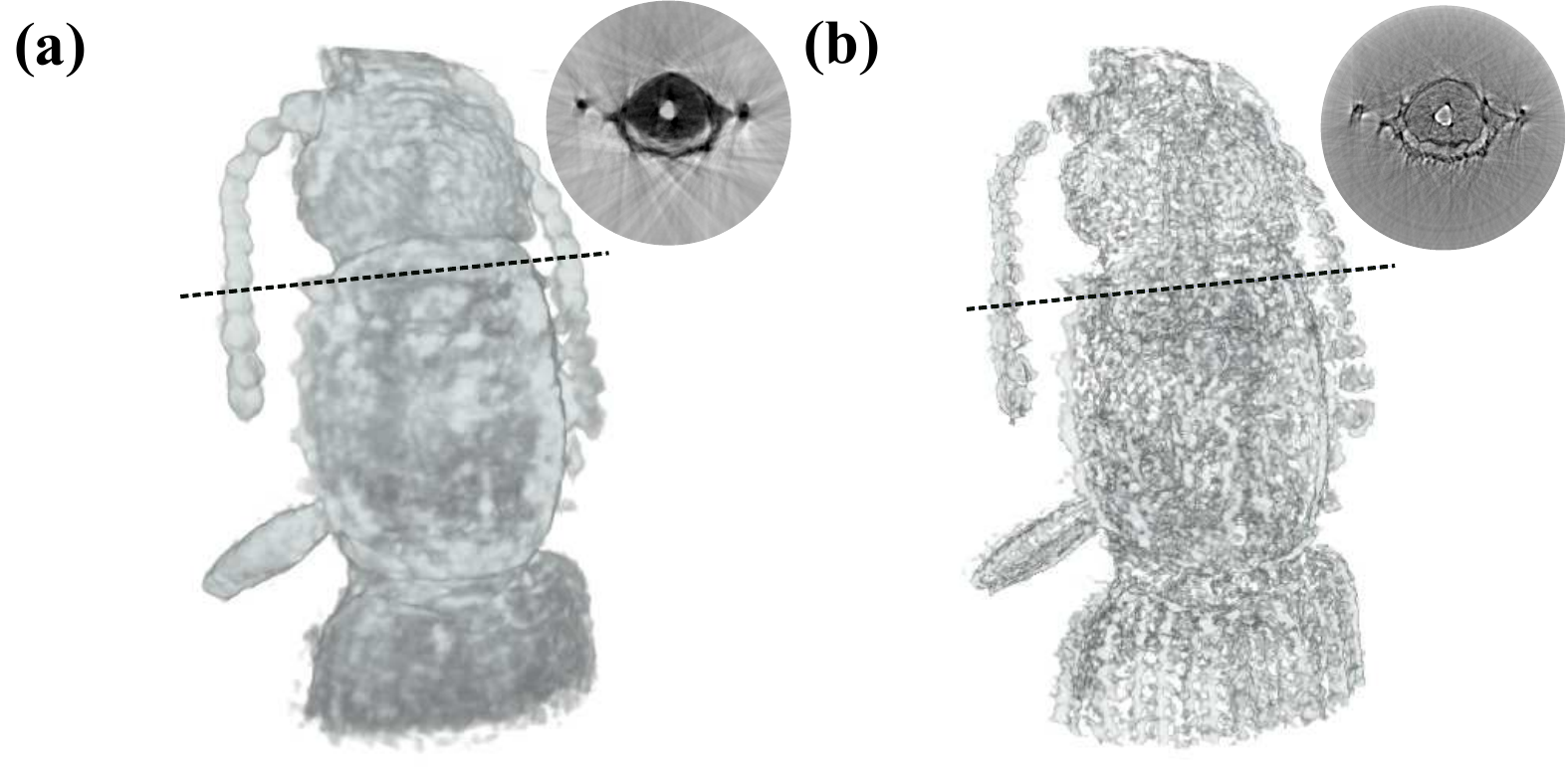}
	\caption{Tomography volume rendering of the 3D (a) phase and (b) transmission of a flour bug. Inserted figures are the line slices.}
	\label{fig:tomo}
\end{figure}

For high-resolution tomography, 1800 projections of the flour bug were acquired with an angular resolution of 0.1\textdegree and an exposure time of 5 seconds. The exposure time is primarily restricted by the X-ray flux of the bending magnet beamline. Using one A100 GPU, the data processing time of each image pair with a size of 2112$\times$2112 pixels was 0.16 seconds. Thanks to the fast data processing speed of SPINNet, the phase and transmission images of the full tomography dataset were obtained within only several minutes. Both the 3D transmission and phase volumes were reconstructed by filtered back projection (FBP) method with a ram-lak filter using the Astra-toolbox~\cite{van2016fast,van2015astra} and shown in Fig.~\ref{fig:tomo}. \textcolor{urlblue}{Visualization 1} and \textcolor{urlblue}{Visualization 2} provide complementary information on the sample. Thanks to the fast speed of SPINNet, the 3D reconstruction can be carried out immediately after the data acquisition, which is beneficial for checking the data quality during the experiment time. The 3D resolution in Fig.~\ref{fig:tomo} is mainly limited by the detector resolution and the angular resolution and stability of the tomography rotation stage. 

\section{Discussion and Conclusion}

SPINNet has been developed for high-resolution real-time phase-contrast imaging and wavefront sensing. SPINNet has an optimized structure featured with multi-resolution analysis and 3D cost volume evaluation. The multi-resolution approach significantly accelerates the reconstruction speed and enables the simultaneous prediction of sample features with different length scales and phase variations. The 3D cost volume is added to the network to provide physical constraints for the network training, which is essential for the prediction accuracy of SPINNet. The 3D cost volume contains the correlation information in addition to the image intensity, which has a similar physical process as the conventional DIC-based analysis.

Thanks to the optimized network structure, SPINNet can be trained with pure simulation data. This characteristic can significantly broaden applications of SPINNet where experimental training data are difficult to acquire. Although the network is trained with simulation data, SPINNet can be applied directly to the actual experimental dataset without any fine-tuning or transfer learning process. We successfully demonstrated applications in quantitative at-wavelength metrology, quantitative phase imaging, and phase-contrast tomography.

SPINNet outperforms the traditional DIC-based XST analysis in terms of the imaging quality and, more significantly, the computation speed. Over two-order of magnitude improvement in speed has been demonstrated compared with the DIC-based analysis, enabling real-time and in-situ measurement for SBI methods. Furthermore, the reconstruction speed of SPINNet can be improved by slightly sacrificing spatial resolution and image quality. As shown in Fig.~S3 in \textcolor{urlblue}{Supplement 1}, a reconstruction speed in the 50~ms level can be achieved for each image pair by using a Net-level 3 model.

In summary, to the best of our knowledge, this is the first time a deep learning method has been applied in the speckle-based imaging field. We anticipate SPINNet will solve the bottleneck of speckle-based imaging techniques in measurement speed. It will pave the way to real-time phase-contrast imaging measurements in a broad range of applications, including real-time X-ray wavefront sensing and in-situ phase-contrast imaging of biomedical samples and soft materials.

\begin{backmatter}
\bmsection{Funding}
U.S. Department of Energy (No. DE-AC02-06CH11357).

\bmsection{Acknowledgement}
This research used resources of the Advanced Photon Source, Argonne Leadership Computing Facility, the Center for Nanoscale Materials, U.S. Department of Energy (DOE) Office of Science User Facilities operated for the DOE Office of Science by Argonne National Laboratory under Contract No. DE-AC02-06CH11357. We gratefully acknowledge the computing resources provided on Swing, a high-performance computing cluster operated by the Laboratory Computing Resource Center at Argonne National Laboratory. ZQ acknowledges partial support from Argonne LDRD 2020-0181 – Advanced Method for Ultrahigh Sensitivity and Resolution At-Wavelength X-ray Optics Characterization.

\bmsection{Disclosures}
The authors declare no conflicts of interest in this paper.

\bmsection{Data Availability}
Data underlying the results presented in this paper are not publicly available but may be obtained from the authors upon reasonable request.

\bmsection{Supplemental document}
See Supplement 1 for supporting content.

\end{backmatter}


\bibliography{references}

\end{document}


\maketitle

\section{Coded phase mask pattern}

\begin{figure}[ht!]
	\centering\includegraphics[width=0.48\textwidth]{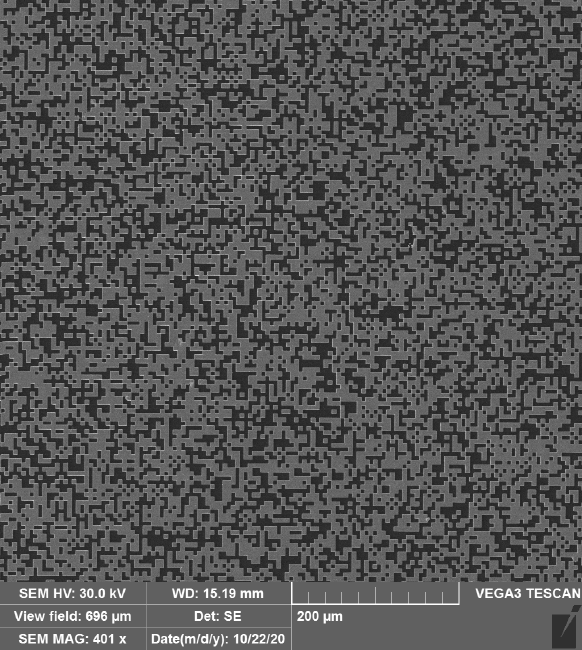}
	\caption{SEM image of a coded binary phase mask with a pitch size of 5~\textmu m.}
  \label{fig:mask_pattern}
\end{figure}

The coded mask was fabricated by electroplating Au (2 \textmu m thick) into polymer molds (poly(methyl methacrylate), PMMA) on a silicon nitride membrane. The bias-corrected coded-mask pattern has a binary pitch size of 5~\textmu m generated using a Python program with a random number generator to create a script for a CNST Nanolithography Toolbox \cite{balram2016nanolithography}. The detailed fabrication process can be found in Ref \cite{marathe2016measurement} with the primary difference being the pattern generation and development described herein. Figure~\ref{fig:mask_pattern} shows the SEM image of the used phase mask. 

\section{Detailed SPINNet structure}

\begin{figure}[ht!]
	\centering\includegraphics[width=1\textwidth]{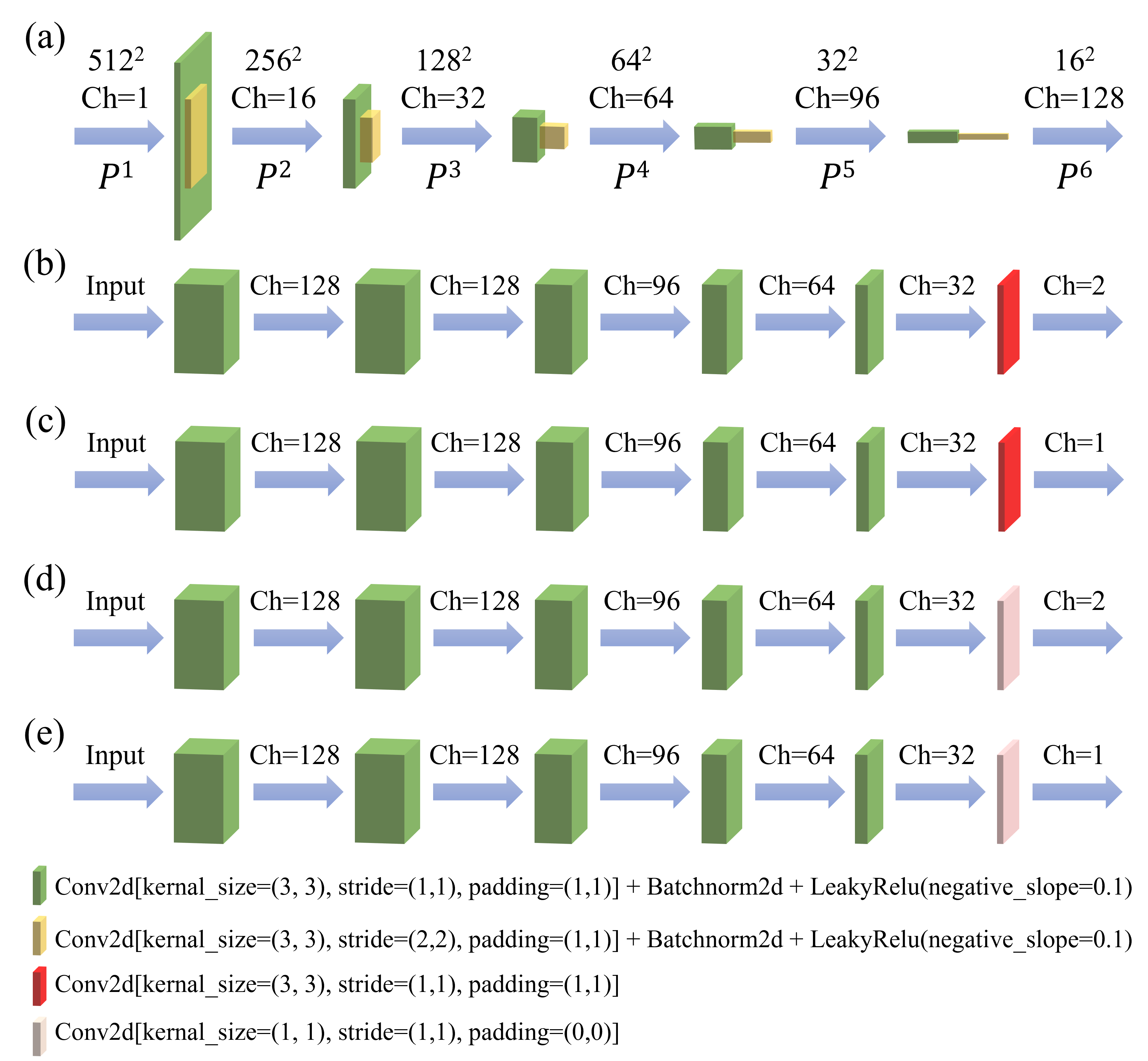}
	\caption{Detailed SPINNet structures for (a) the feature extractor, (b) the phase estimator (PhaseNet), (c) the transmission estimator (TNet), (d) the phase refiner (PhaseRefiner), and (e) the transmission refiner (TRefiner).}
  \label{fig:detail_network}
\end{figure}

The detailed structures of SPINNet feature extractor and phase and transmission estimators and refiners are shown in Fig.~\ref{fig:detail_network}. The blocks represent series of neural network operations. 

The feature extractor has a similar structure as the encoder of the well known encoder-decoder neural network structure. The green block in Fig.~\ref{fig:detail_network}(a) represents a fundamental procedure consisting of a 2D convolution (Conv2d) layer with a stride of 1, a 2D batch normalization (Batchnorm2d) layer, and a leaky Relu nonlinear activation (LeakyRelu) layer. The yellow block includes a Conv2d layer with a stride of 2, a Batchnorm2d layer, and a LeakyRelu layer. Because of the different stride settings in the Conv2d layer, the yellow block can down-sample the input data, as shown in Fig.~\ref{fig:detail_network}(a). The input simulation data for training has 512$\times$512 pixels with a single channel, which will be down-sampled by a factor of 2$\times$2 but with an increasing channel number from 1 to 128. The feature extractor networks for the reference and sample images have the same structure but are independent.

Figures~\ref{fig:detail_network}(b) and (c) are the detailed network structures of the phase and transmission estimators (PhaseNet and TNet), respectively. Both estimator networks consist of five consequent green blocks (Conv2d+Batchnorm2d+LeakyRelu) and a final Conv2d block with a kernel size of 3 [red blocks in Figs.~\ref{fig:detail_network}(b) and (c)]. The only difference between the phase and transmission estimators is the final output layer with an output channel of 2 and 1, respectively. 

The phase and transmission refiners (PhaseRefiner and TRefiner) are shown in Fig.~\ref{fig:detail_network}(d) and (e), respectively. Their structures are similar to the estimators, except the final Conv2d layer has a kernel size of 1 instead of 3. Test results showed that the kernel size of 1 can provide a higher spatial resolution because of a weaker low-pass filtering effect than a kernel size of 3. The difference between the phase and transmission refiners is again on the different output channel numbers.

\section{Prediction comparison with different network levels}

\begin{figure}[ht!]
	\centering\includegraphics[width=0.98\textwidth]{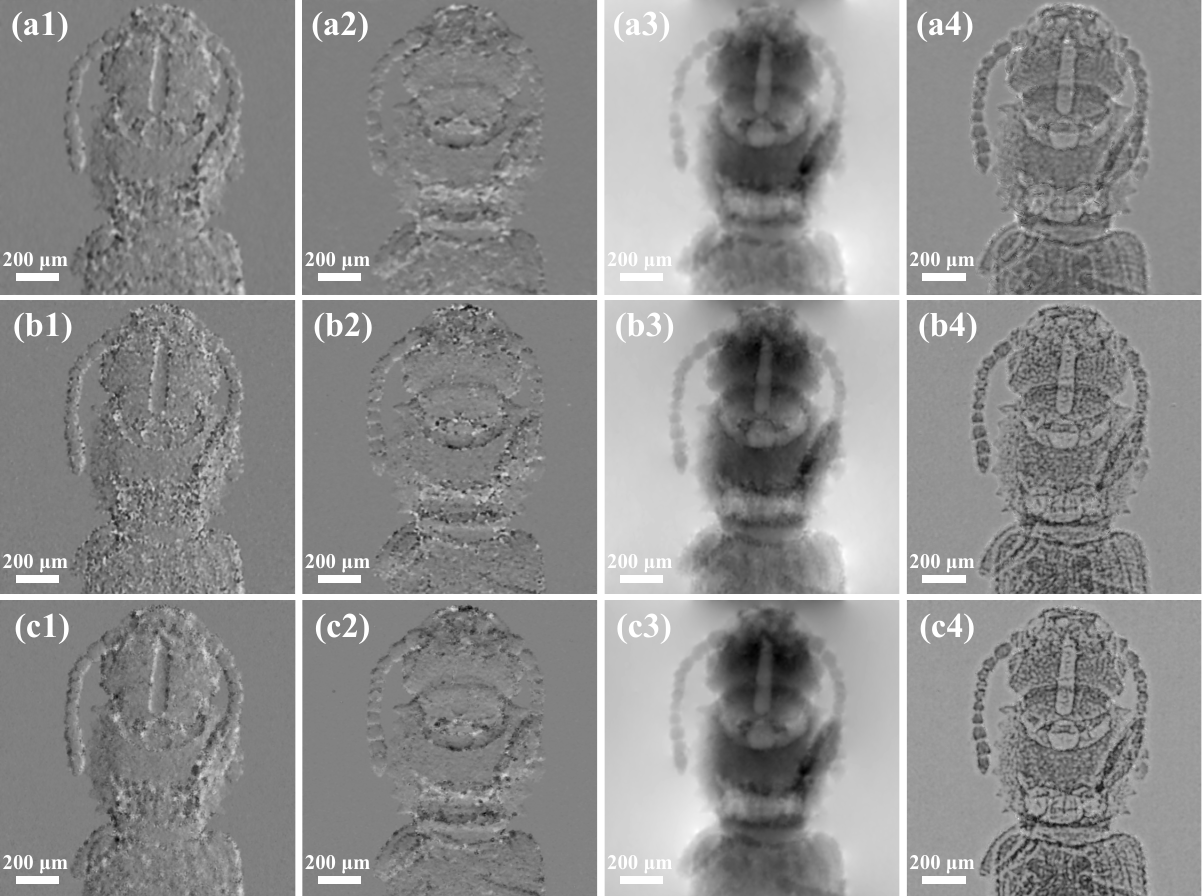}
	\caption{SPINNet predictions of (1) the horizontal differential phase, (2) the vertical differential phase, (3) the phase image, and (4) the transmission image of a flour bug using (a) stage 1 model with Net-level 6\textasciitilde3, (b) stage 2 model with Net-level 6\textasciitilde2, and (c) stage 3 model with Net-level 6\textasciitilde2 and the refiners.}
  \label{fig:multi_level}
\end{figure}

SPINNet was trained with three stages: stage 1, Net-level 6\textasciitilde3; stage 2, Net-level 6\textasciitilde2; and stage 3, Net-level 6\textasciitilde2 with refiners. Figure~\ref{fig:multi_level} shows the predicted differential phase, phase, and transmission images using the trained models of stage 1 [Fig.~\ref{fig:multi_level}(a)], stage 2 [Fig.~\ref{fig:multi_level}(b)], and stage 3 [Fig.~\ref{fig:multi_level}(c)]. The sample boundaries in the predicted phase and transmission images using the stage 1 model are less sharp and noisier than the prediction of stage 2 and stage 3 models. By adding the phase and transmission refiner, the stage 3 model further improved the imaging quality. The prediction time using the stage 1, stage 2, and stage 3 models for a full-field image pair is 72~ms, 125~ms, and 160~ms with a batch size of 1 (50~ms, 100~ms, and 130~ms with a batch size of 3), respectively. Considering the already good imaging quality of the stage 1 model as shown in Fig.~\ref{fig:multi_level}(a), the SPINNet efficiency can be pushed to a 50~ms level with slightly sacrificing image quality. This can be potentially beneficial for applications requiring faster speed than better imaging quality, such as a real-time wavefront sensor in a fast feedback beamline control system.

\section{Prediction results for different coded masks}

\begin{figure}[ht!]
	\centering\includegraphics[width=0.78\textwidth]{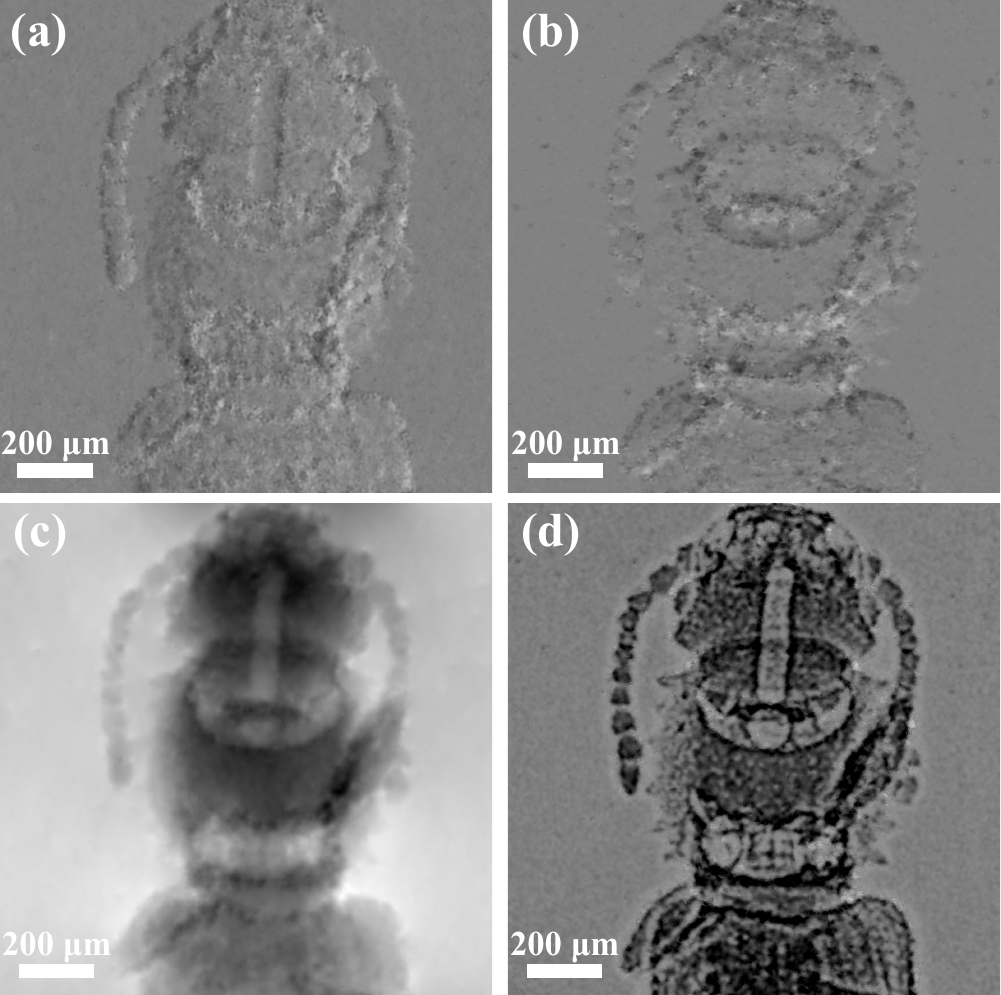}
	\caption{SPINNet predictions of (a) the horizontal differential phase, (b) the vertical differential phase, (c) the phase image, and (d) the transmission image of a flour bug. The SPINNet model was trained with simulated data using coded mask patterns with 5~\textmu m pitch, while the experimental data was obtained using a coded mask with 2~\textmu m pitch.}
  \label{fig:mask_2um}
\end{figure}

All simulated training data for SPINNet was generated based on different coded mask patterns with a pitch size of 5~\textmu m, and it has been proved that the trained model worked on the experimental data using a coded mask with the same pitch size, as shown in Fig.~\ref{fig:multi_level}. Here, we demonstrate that the same SPINNet model also works on data with different coded mask parameters. Figure ~\ref{fig:mask_2um} shows the SPINNet predicted differential phase, phase, and transmission images of the same flour bug but with experimental data using a coded mask of 2~\textmu m pitch. Compared with the results in Fig.~\ref{fig:multi_level} (c1\textasciitilde c4), images in Fig.~\ref{fig:mask_2um} have only slightly higher noise. Even though the trained SPINNet model has never seen a 2~\textmu m speckle pattern, the predicted results still show high image quality and resolution.



